**Unusual ultra-low frequency fluctuations in freestanding graphene**

P. Xu[1], M. Neek-Amal[2,#], S. D. Barber[1], M. L. Ackerman[1], J. K. Schoelz[1], P. M. Thibado[1]*, A. Sadeghi[3] and F. M. Peeters[2]

[1]*Department of Physics, University of Arkansas, Fayetteville, Arkansas 72701, USA*

[2]*Departement Fysica, Universiteit Antwerpen, Groenenborgerlaan 171, B-2020 Antwerpen, Belgium*

[#]*Department of Physics, Shahid Rajaee Teacher Training University, Lavizan, Tehran 16788, Iran*

[3]*Departement Physik, Universat Basel, Klingelbergstrasse 82, CH-4056 Basel, Switzerland*

*Correspondence and requests for materials should be addressed to P.M.T. (e-mail: thibado@uark.edu).

**Abstract**

Intrinsic ripples in freestanding graphene have been exceedingly difficult to study. Individual ripple geometry was recently imaged using scanning tunneling microscopy, but these measurements are limited to static configurations. Thermally-activated flexural phonon modes should generate dynamic changes in curvature. Here we show how to track the vertical movement of a one-square-angstrom region of freestanding graphene using scanning tunneling microscopy, thereby allowing measurement of the out-of-plane time trajectory and fluctuations over long time periods. We also present a model from elasticity theory to explain the very-low frequency oscillations. Unexpectedly, we sometimes detect a sudden colossal jump, which we interpret as due to mirror buckling. This innovative technique provides a much needed atomic-scale probe for the time-dependent behaviors of intrinsic ripples. The discovery of this novel progenitor represents a fundamental advance in the use of scanning tunneling microscopy which together with the application of a thermal load provides a low frequency nano-resonator.

**INTRODUCTION**

Tracking the movement of individual atoms has its origins in field ion microscopy[1]. The invention of scanning tunneling microscopy (STM), however, marked the birth of a new era, famously allowing Mo *et al.*, for instance, to observe the migration of Si atoms on the Si(001) surface and extract the activation energy for diffusion[2]. Swartzentruber went a step further by programming the STM tip to physically follow each step of a single diffusing atom, thereby recording its site-to-site movement[3]. In this tradition, we utilize STM to precisely monitor the out-of-plane motions of a one-square-angstrom region in freestanding graphene for the first time. These fluctuations[4] are linked to the flexural phonon modes, which are critical to many of graphene's amazing properties, such as its previously unpredicted stability[5, 6], anomalous negative coefficient of thermal expansion[7], and efficient thermal conductivity[8, 9]. The ripples which form[10, 11] are also important for understanding transport properties in graphene, as they decrease the carrier mobility in freestanding membranes[12], cause disruptive charge puddling[13], and induce strong pseudo-magnetic fields[14]. The ability to locally measure the thermal fluctuations as a function of time will open the door to directly testing their role in these and other important processes.

## RESULTS

**STM tip interaction with freestanding graphene.** Increasingly, STM is becoming the tool of choice to manipulate and map suspended graphene[15-18]. As the STM tip approaches a roughened, freestanding graphene layer, depicted in Fig. 1a, it may have two significant effects on the sample. (For full system and sample descriptions, see Methods and Supplementary Fig. 1.) The first depends on the potential difference $V$ between the positively biased tip and the grounded sample, as demonstrated in the inset of Fig. 1b. When the voltage is increased from 0.1 V to 3.0 V at a constant tunneling current of 1.00 nA, the tip height increases 20 nm. From the size of this displacement and the requirement of constant current, we can conclude that the flexible graphene has also moved about 20 nm locally due to electrostatic attraction[19]. Since the electrostatic force depends quadratically[20] on the bias voltage ($\propto V^2$), we re-plot $V^2$ as a function of tip height in the main graph of Fig. 1b. The main plot is divided into four regions for discussion purposes: (I) low-bias voltages where $V^2 < 0.1$ V$^2$, (II) voltages between 0.1 V$^2$ and 1 V$^2$, (III) voltages between 1 V$^2$ and 4 V$^2$, and (IV) voltages beyond 4 V$^2$. In region (I) where the film has not been pulled very high, we believe the thermal contribution to the energy is much larger than the electrostatic one. After that, in region (II) where most of the height change happens, the electrostatic force ($F_e \propto V^2$) is proportional to the height variation ($\Delta Z$), and the system obeys Hooke's law (linear region). Next, in region (III) the electrostatic force is no longer proportional to the height variation, and the nonlinear effects become important. Finally, in region (IV) there is no further increase in the height with increasing voltage, meaning the graphene sheet is fully pulled up (i.e., tense membrane boundary condition) and becomes inflexible.

The second important effect caused by the STM tip is heating of the sample by passing current through it, as evidenced by Fig. 1c. Here the tunneling current is ramped from 0.01 nA to 1.00 nA at a constant tip bias of 0.1 V, and the displacement is −20 nm. Note this behavior cannot be due to an electrostatic effect. The tunneling gap will be slightly reduced in order to accommodate the slightly larger setpoint current, but this would increase the electrostatic force, resulting in a height increase rather than a decrease. However, increasing the current does locally heat the graphene[21]. Thus this behavior can be attributed to the negative thermal expansion coefficient of graphene, which causes it to contract as the temperature of the graphene under the tip is increased. We assume that in the steady state, the temperature rise under the tip is proportional to the power of the Joule heating at the junction ($\propto I^2$). Then the local height of the graphene under the tip decreases as $h = h_0 e^{-kI^2}$ due to contraction, where $h_0 = h(I = 0)$ depends on $V$, while the constant $k$ depends on graphene's thermal conductivity and expansion coefficient, as well as on the tip-sample separation $D = \frac{1}{2c}\ln(V/R_0 I)$ (see Supplementary Note 1). A typical value of the decay constant $c$ is 1 Å$^{-1}$, and the contact resistance $R_0$ is typically 13 kΩ. However, at constant $V$ we can let $D = a \ln(I)$ and determine the unknown parameters by fitting the total tip height

$$\text{Height}(I) = h_0 e^{-kI^2} + a\ln(I) + z_{\text{offset}} \qquad (1)$$

to the experimental data, as shown by the red curve in Fig. 1c, which is in good agreement with our experimental results. Note that the offset is necessary because the absolute starting height is unknown; it is set to zero for convenience. Although our fluctuation measurements take place at a constant voltage $V_0$ and constant current $I_0$, the two simple experiments above clearly indicate that the chosen tunneling parameters may have a significant impact on our results.

**Fluctuation behaviors of freestanding graphene and theoretical analysis.** Three different qualitative categories of behavior can be distinguished among the collected fluctuation information. We call the first and most common type, featured in Fig. 2, "random" because no pattern or special feature is apparent. All data was recorded by taking a 0.1 nm × 0.1 nm STM image on freestanding graphene, such as the example shown in Fig. 2a (lower image). To understand the size of this scan relative to the lattice of atoms in graphene, refer to the 0.1 nm × 0.1 nm square outlined in the upper STM image, which was acquired separately from a stabilized region of the same freestanding graphene sample. The lower image (400 × 400 data points) shows the height of the tip in a black-to-orange-to-white color scale and is oriented such that the slow scan direction is horizontal and the fast scan is vertical. Note that significant changes in height rarely occur within a single vertical line. Therefore, a height-time signal was extracted along the slow scan direction by averaging the 400 points in each vertical line, resulting in a sampling frequency of 1 Hz over 400 s, as shown in Fig. 2b. The unique power of this technique is that it effectively monitors the vertical position of a one-atom sized region for a long period of time. Normal imaging covers too large an area to determine how any one location changes on a short timescale, and spectroscopy measurements like those in Fig. 1 happen too quickly to observe these fluctuations, though focused in one spot.

The result on freestanding graphene is that the height displacements greatly exceed the scale of the scan area. In this low bias regime (0.01 V), thermal energy is dominant[22]. The graphene height in Fig. 2b fluctuates around zero (average height was subtracted from the raw data) with a standard deviation of 1.47 nm, an indication of the amplitude of wrinkles present in graphene (see Supplementary Fig. 2). For comparison, data taken in exactly the same manner on a sample of gold is also plotted, showing only the smallest changes in height over the entire scan. In addition, we always verify that graphene is indeed moving with the tip by measuring the tunneling current as a function of time, displayed in Fig. 2c. It is almost constant at the setpoint value (0.2 nA), and the variations that do occur are not large enough to account for 3 nm fluctuations in the tip height without the sample also moving. We further characterize each height-time data set by calculating its autocovariance function $A(t)$, which is plotted in Fig. 2d. $A(t)$ decays exponentially for random fluctuations, indicating that they are only correlated for short periods of time. In this example, the decay constant is 8 s.

When we increase the tunneling current, freestanding graphene exhibits noticeable periodic oscillations as represented in Fig. 3. Two samples of height-time data, taken at different currents but at the same location and the same low voltage (to avoid stretching), are displayed in Fig. 3a. The most noticeable difference is in the peak-to-peak amplitude; however, the fluctuations in both appear to have a roughly repeated character, which was confirmed by calculating $A(t)$ for each curve, plotted in Fig. 3b. Its peaks share the periodicity of the signal, which is clearly around 100 s at 3 nA and 50 s at 5 nA. Conveniently, the linear susceptibility $\chi(t)$ can also be obtained from $A(t)$ according to

$$\chi(t) = -\frac{1}{k_B T}\frac{dA(t)}{dt} \tag{2}$$

where $k_B$ is the Boltzmann constant, and $T$ is the absolute temperature[23]. $dA/dt$ is therefore plotted as an inset for each curve in Fig. 3b and gives a maximum susceptibility of 5 nm²/s. Furthermore, $\hat{\chi}(\omega)$, the Fourier transform of $\chi(t)$, can in turn be related to the power spectral density $S(\omega)$ by the fluctuation-dissipation theorem, which states

$$S(\omega) = \frac{2k_B T}{\omega} \text{Im}[\hat{\chi}(\omega)] \tag{3}$$

where $\omega$ is the frequency. The periodogram, an estimate of $S(\omega)$, was computed for the periodic results and plotted in Fig. 3c. The 3 nA spectrum has a prominent maximum near a frequency of 0.01 Hz, and gives a maximum power spectral density of 9,000 nm²/Hz. Meanwhile, the 5 nA spectrum peaks around 0.02 Hz, and gives a smaller maximum power spectral density of 1,500 nm²/Hz. These low frequency peaks represent the dissipation of thermal energy through flexural acoustic modes near the Brillouin zone center[24-26].

However, if we model the freestanding graphene sample as a doubly-clamped resonator with length $L = 7.5$ µm (see Methods) and mass density $\rho$ subjected to an initial strain $\varepsilon$, plate theory predicts a resonance frequency given by $f = \frac{1}{2L}\sqrt{\frac{Y}{\rho}\varepsilon}$, where $Y = 340$ N/m is the Young's modulus. This results in frequencies in the GHz range[27], which is much too large. Therefore, in order to explain the experimental data using elasticity theory, we must invoke a different mechanism. Since the measurements in Fig. 3a are performed in the limit of high electric current (i.e. temperature) and low bias voltage (but it is not negligible, see Supplementary Note 2), there are additional terms in the stress tensor[28]:

$$\sigma_{ij} = 2\mu\left(\varepsilon_{ij} - \frac{1}{2}\delta_{ij}\sum_i \varepsilon_{ii}\right) + B\left(\sum_i \varepsilon_{ii} + \alpha\Delta T\right)\delta_{ij} \qquad i,j = 1,2 \tag{4}$$

where the Lamé coefficient $\mu$ characterizes the shear rigidity of graphene, $B = \frac{Y\mu}{4\mu-Y}$ is the bulk modulus [29], and $\alpha \approx 5 \times 10^{-6}$ K$^{-1}$ is the absolute value of the thermal expansion coefficient at room temperature. The last term represents the thermal stress due to the temperature gradient near the tip region and the supports. In the steady state, $\Delta T$ is constant for a given electric current. In the presence of thermal stress, using equation (4) and the corresponding equation of motion for the graphene membrane results in the following flexural phonon frequency (see Supplementary Note 3):

$$\omega_q = q\sqrt{\frac{\kappa q^2 + \tau}{\rho}} \qquad (5)$$

where $\kappa$ is the bending rigidity of graphene, $\rho \cong 7.6 \times 10^{-7}$ kg m$^{-2}$ is the mass density, and $\tau = -B\alpha\Delta T$ is the effective negative surface tension. Since $\Delta T > 0$, equation (5) can result in low frequencies. The critical wave vector is determined as

$$q_c(\text{Å}^{-1}) = \sqrt{\frac{|\tau|}{\kappa}} \qquad (6)$$

Using typical values[12] of $\kappa = 1.1$ eV and $B = 208$ N/m, we can estimate the critical wave vector, which attains realistic values for low $\Delta T$. For example, at $\Delta T \simeq 10$ K, $\Delta T \simeq 20$ K, and $\Delta T \simeq 100$ K the corresponding wavelengths are around 26 nm, 18 nm, and 8 nm, respectively. Note that longer critical wavelengths would be expected when other external stress sources exist; for example, when scanning close to the boundary or in the presence of a strong asymmetry.

By defining $\gamma = \frac{q}{q_c}$, we can rewrite equation (5) as

$$\omega_q = \gamma\sqrt{(\gamma^2 - 1)\frac{\kappa}{\rho} q_c^2} \qquad (7)$$

And for $q$ near to $q_c$, equation (7) results in small frequencies and $\omega_{q_c} \propto \Delta T \propto I^2$. Therefore, the larger the electric current, the larger the frequency when $q$ is very close to $q_c$. We confirm this prediction in Fig. 3d by plotting the primary oscillation frequency as a function of $I^2$ and a linear trend line for four different scans obtained at the same low bias voltage. Furthermore, we can deduce based on the aforementioned theory that the bottom panel in Fig. 3a is more affected by thermal stress than the top panel. Also, the height variance should approximately obey $\langle h^2 \rangle \propto I^{-4}$, which is in agreement with these four different electric currents as shown in Fig. 3e. In fact, these results are evidence for the softening of flexural phonons in the case of compressive strain in graphene. This result has been theoretically predicted using density functional theory calculations combined with non-linear classical elasticity theory[7, 30].

The last category of fluctuations discovered appears to exhibit a "mirror buckling" effect and is explored in Fig 4. The example STM image shown in Fig. 4a is very dark for the first fourth of the image and nearly white for the rest, signaling that a sudden jump has occurred which is much larger than the variations before or after. A height-time profile, extracted in the manner previously described, is plotted in Fig. 4b to reveal that the membrane surprisingly undergoes an enormous 60 nm displacement around $t = 100$ s but oscillates relatively little after the jump, somewhat similar to a critical transition. Examining the tunneling current as a function of time as provided in Fig. 4c, verifies that the sudden displacement is a real effect. Because the current is reasonably constant at 4 nA except for a spike coinciding with the jump in height, and since the spike is well below the system's saturation current of 50 nA, tunneling was never compromised. Height profiles such as this yield an unusual $A(t)$, plotted in Fig. 4d. It takes a much longer time to decay, and its value at $t = 0$ (equivalent to the variance $\langle h^2 \rangle$) is anomalously large. The snap-through behavior can typically be produced by performing a series of scans at progressively larger currents, keeping the voltage constant. The Fig. 4 data was part of such a series, and the variance from each successive scan is recorded in Fig. 4e. The large peak at 4 nA corresponds to the critical-like transition, which normally only occurs once as the current is increased within these limits, but not always at the same current.

This unusual and unexpected event is a result of heating the sample in the presence of an attractive force. We can interpret this effect as due to mirror buckling of a thin shell. For example, for a spherical shell with radius $R$, the critical pressure which causes mirror buckling is estimated as[31, 32]

$$p_c \propto \sqrt{\frac{\kappa Y}{R^4}} \qquad (8)$$

where $p_c \propto V_c^2$, and both $\kappa$ and $Y$ will be modified by the presence of heating and an electrostatic force. By increasing temperature (electric current), $Y$ is reduced, and a lower bias voltage is required to invert the buckled structure. Also, a larger radius lowers the voltage that leads to a flip. At this specific current and voltage, a large buckled region below the tip is heated for 100 s and then suddenly changes its curvature from negative to positive, i.e., it switches from a bowl-like to a bump-like shape. This process is illustrated in Fig. 4f. During Stage 1 (also labeled in Fig. 4b) the height increases as the tip begins to deform the bowl. A metastable state is reached at Stage 2 where the force is not large enough to keep altering the local graphene configuration. However, the critical pressure is finally reached as the temperature continues to rise, leading the system to suddenly switch to the more stable bump shape shown in Stage 3.

The ripples of freestanding graphene were first imaged using STM in the pioneering work of Zan et al.[15]. They were able to establish the general topography of the ripples, which is the critical link to the general stability of 2D systems. What they found, using a STM setup similar to ours, was that the height of the ripples is on the order of 1 nm and that the wavelength is on the order of 5 to 10 nm. The researchers also report that the ripple structure is static in time and is stable enough to repeatedly image the surface topography. At first, this seems to contradict our results. However, Zan et al. also report that the freestanding graphene sheet has regions that are too unstable to image with the STM (i.e., when they look in a region far from the copper bar supports). This is also what we found. The surface of the graphene fluctuates too quickly for the STM to acquire constant-current images of the surface topography. However, we have been able to tunnel into the graphene sheet at that one spot, maintain a constant current, and record the height changes occurring in time. Sometimes the surface will spontaneously stabilize as shown in Fig. 4b, and then, if it stays that way, it can be imaged with the STM. We also found that the surface can be locally stabilized using constant current height-voltage sweeps up to 3 V (unpublished).

**DISCUSSION**

We have here demonstrated a way to gain unprecedented insight into the dynamics of ripples of freestanding graphene using the versatility of STM. While imaging an area corresponding to a single carbon atom, the vertical movement of the graphene is easily monitored with unparalleled precision. A degree of control over the shape and temperature of the membrane is applied through changing the tunneling parameters. The observed fluctuations are generally large and can be sorted based on exhibiting random, periodic, or mirror-buckling behavior. In particular, the periodic oscillations and their current-dependent characteristics were shown to be consistent with the predictions of elasticity theory under the influence of thermal stress. No other technique has demonstrated the ability to probe such low-frequency flexural phonon modes at the atomic scale, permitting direct investigation of the dynamic ripples that affect almost every property of graphene. Furthermore, the biological time-scale of the fluctuations (50 s and 100 s) observed here advanced our fundamental understanding of membrane dynamics. We expect that the new technique will lead to new experiments in graphene, ranging from fundamental to advanced quantum charge control and thermal load application.

**Methods**

**Scanning tunneling microscopy.** An Omicron STM was used to obtain constant-current STM images of gold and freestanding graphene, as well as feedback-on measurements of tip height as a function of either setpoint current or bias voltage. The tip points upward. All images were collected at a scale of 0.1 nm × 0.1 nm, at a scan rate of 0.2 nm/s, and had 400 data points per line with 400 lines per image. For each line, the topographical data was collected simultaneously with the tunneling current in both the forward and backward scanning directions. In this way each line took one second to complete for a total of 400 seconds per image. All STM images presented are 400 × 400 pixels and have a black-to-orange-to-white color scale.

The electronics (cabling and pre-amplification in particular) used for these experiments effectively act as a low-pass filter with a bandwidth (-3 dB) of approximately 800 Hz. In addition, the speed at which we acquire digital data is 800 points per second, so we have a maximum frequency of 800 Hz that we can measure. Finally, for the height-time data presented in Figs. 2, 3, and 4 we collect digital data for 400 s, giving us a minimum detectable frequency of 0.0025 Hz. The error bars added to Figs. 1b and 1c represent the standard deviation of ten repeated measurements taken consecutively at the same location of the sample. The error bars added to Figs. 2b, 3a, and 4b represent the standard deviation of the 400 data points in the fast scan direction of the acquired 400 by 400 data point STM image. The results presented here are typical examples. We have collected numerous data sets from each location we studied, we have studied numerous locations on each sample, and we have studied many different samples.

The graphene was grown using chemical vapor deposition[33], then transferred onto a 2000-mesh, ultrafine copper grid—a lattice of square holes 7.5 μm wide and bar supports 5 μm wide—by the commercial provider. Grid was mounted on a flat tantalum sample plate in the STM chamber and grounded. Data was acquired using tungsten tips made in-house. A positive bias voltage was applied to the tip in all experiments.

These are difficult STM measurements, so we are adding some extra details here about our effort. First, the freestanding graphene is attached to only one side of the Cu grid. From our scanning electron microscope and energy dispersive X-ray analysis experiments (not reported here), we have confirmed that the freestanding graphene covers 90% of the surface and is free of detectable contamination. We have successfully done the experiments reported here with the freestanding graphene side of the sample facing up as well as facing down. When the sample is facing down, the Cu grid is supported on a stand-off support so the graphene cannot come into contact with the STM sample holder. The STM tip monitors the freestanding graphene through the holes of the Cu grid, and we have had fewer problems with stability in the downward-facing configuration. When the sample is loaded into the STM load-lock chamber, only lamp-type heating is applied, and no other heating or annealing is done to the sample. We are concerned that excessive annealing may cause macroscopic texturing due to the Cu grid expanding and the graphene contracting during the heating process[11]. About 10% of the time, after approaching the freestanding graphene surface, the STM cannot establish stable constant-current tunneling and we move to a new location. Most of the time, we can establish stable constant-current tunneling, but once we attempt to acquire a STM image of any appreciable size the constant-current tunneling cannot be maintained. Consequently, we reduced the scan size to a single spot and then recorded the movement of the surface in time while maintaining a constant tunneling current.

**References**

1. Ehrlich, G. & Hudda, F.G. Atomic view of surface self-diffusion: tungsten on tungsten. *J. Chem. Phys.* **44**, 1039-1049 (1966).

2. Mo, Y.W., Kleiner, J., Webb, M.B. & Lagally, M.G. Activation energy for surface diffusion of Si on Si(001): a scanning-tunneling-microscopy study. *Phys. Rev. Lett.* **66**, 1998-2001 (1991).

3. Swartzentruber, B.S. Direct measurement of surface diffusion using atom-tracking scanning tunneling microscopy. *Phys. Rev. Lett.* **76**, 459-462 (1996).

4. Smolyanitsky, A. & Tewary, V.K. Manipulation of graphene's dynamic ripples by local harmonic out-of-plane excitation. *Nanotechnology* **24**, 055701 (2013).

5. Fasolino, A., Los, J.H. & Katsnelson, M.I. Intrinsic ripples in graphene. *Nature Mater.* **6**, 858-861 (2007).

6. Meyer, J.C. *et al.* On the roughness of single- and bi-layer graphene membranes. *Solid State Commun.* **143**, 101-109 (2007).

7. de Andres, P.L., Guinea, F. & Katsnelson, M.I. Bending modes, anharmonic effects, and thermal expansion coefficient in single-layer and multilayer graphene. *Phys. Rev. B* **86**, 144103 (2012).

8. Lindsay, L., Broido, D.A. & Mingo, N. Flexural phonons and thermal transport in graphene. *Phys. Rev. B* **82**, 115427 (2010).

9. Balandin, A.A. *et al.* Superior thermal conductivity of single-layer graphene. *Nano Lett.* **8**, 902-907 (2008).

10. Meyer, J.C. *et al.* The structure of suspended graphene sheets. *Nature* **446**, 60-63 (2007).

11. Bao, W. *et al.* Controlled ripple texturing of suspended graphene and ultrathin graphite membranes. *Nature Nanotech.* **4**, 562-566 (2009).

12. Castro, E.V. *et al.* Limits on charge carrier mobility in suspended graphene due to flexural phonons. *Phys. Rev. Lett.* **105**, 266601 (2010).

13. Partovi-Azar, P., Nafari, N. & Tabar, M.R.R. Interplay between geometrical structure and electronic properties in rippled free-standing graphene. *Phys. Rev. B* **83**, 165434 (2011).

14. Castro Neto, A.H., Guinea, F., Peres, N.M.R., Novoselov, K.S. & Geim, A.K. The electronic properties of graphene. *Rev. Mod. Phys.* **81**, 109-162 (2009).

15. Zan, R. *et al.* Scanning tunnelling microscopy of suspended graphene. *Nanoscale* **4**, 3065-3068 (2012).

16. Eder, F.R. *et al.* Probing from both sides: reshaping the graphene landscape via face-to-face dual-probe microscopy. *Nano Lett.* **13**, 1934−1940 (2013).

17. Klimov, N.N. *et al.* Electromechanical properties of graphene drumheads. *Science* **336**, 1557-1561 (2012).

18. Mashoff, T. *et al.* Bistability and oscillatory motion of natural nanomembranes appearing within monolayer graphene on silicon dioxide. *Nano Lett.* **10**, 461-465 (2010).

19. Xu, P. *et al.* Atomic control of strain in freestanding graphene. *Phys. Rev. B* **85**, 121406(R) (2012).

20. Sadeghi, A. *et al.* Multiscale approach for simulations of Kelvin probe force microscopy with atomic resolution. *Phys. Rev. B* **86**, 075407 (2012).

21. Flores, F., Echenique, P.M. & Ritchie, R.H. Energy dissipation processes in scanning tunneling microscopy. *Phys. Rev. B* **34**, 2899-2902 (1986).

22. Los, J.H., Katsnelson, M.I., Yazyev, O.V., Zakharchenko, K.V. & Fasolino, A. Scaling properties of flexible membranes from atomistic simulations: application to graphene. *Phys. Rev. B* **80**, 121405(R) (2009).

23. Marconi, U.M.B., Puglisi, A., Rondoni, L. & Vulpiani, A. Fluctuation-dissipation: response theory in statistical physics. *Phys. Rep.* **461**, 111-195 (2008).

24. Seol, J.H. *et al.* Two-dimensional phonon transport in supported graphene. *Science* **328**, 213-216 (2010).

25. Mariani, E. & von Oppen, F. Flexural phonons in free-standing graphene. *Phys. Rev. Lett.* **100**, 076801 (2008).

26. Bonini, N., Lazzeri, M., Marzari, N. & Mauri, F. Phonon anharmonicities in graphite and graphene. *Phys. Rev. Lett.* **99**, 176802 (2007).

27. Chen, C. Graphene NanoElectroMechanical Resonators and Oscillators, Ph.D. Thesis (Columbia University, New York, NY, 2013).

28. Landau, L.D. & Lifshitz, E.M. Theory of Elasticity (Pergamon Press, Oxford, 1970).

29. Zakharchenko, K.V., Katsnelson, M.I. & Fasolino, A. Finite temperature lattice properties of graphene beyond the quasiharmonic approximation. *Phys. Rev. Lett.* **102**, 046808 (2009).

30. de Andres, P.L., Guinea, F. & Katsnelson, M.I. Density functional theory analysis of flexural modes, elastic constants, and corrugations in strained graphene. *Phys. Rev. B* **86**, 245409 (2012).

31. Pogorelov, A.V. Bendings of Surfaces and Stability of Shells (American Mathematical Society, Providence, RI, 1988).

32. Lindahl, N. *et al.* Determination of the bending rigidity of graphene via electrostatic actuation of buckled membranes. *Nano Lett.* **12**, 3526-3531 (2012).

33. Reina, A. *et al.* Large area, few-layer graphene films on arbitrary substrates by chemical vapor deposition. *Nano Lett.* **9**, 30-35 (2009).

**Figure captions**

**Figure 1 | Interaction between STM tip and freestanding graphene. a**, Depiction of a biased STM tip as it is brought near a suspended, grounded graphene sample with intrinsic roughening. As in our STM system, the tip approaches from below, but the measured tip height increases as it moves away from the sample. **b**, The change in STM tip height at one location on freestanding graphene as a function of bias voltage. It increases by 20 nm as the voltage is ramped from 0.1 V to 3.0 V at 1.00 nA, as shown in the inset. The main plot is rescaled to show $V^2$ as a function of height because $V^2$ is proportional to the electrostatic force. The four regimes present in the data

are: I) thermal fluctuations, II) linear elasticity, III) nonlinear elasticity, and IV) boundary effects. **c**, The change in STM tip height at one location on freestanding graphene as a function of tunneling current. The experimental height decreases by 20 nm as the current is raised from 0.01 nA to 1.00 nA at 0.1 V, and this behavior can be replicated by equation (1).

**Figure 2 | Example of random thermal fluctuations. a**, (top) A 1 nm × 1 nm filled-state STM image of freestanding graphene with a box highlighting the size of a one-square-angstrom region. (bottom) A one-square-angstrom STM image taken at constant current (0.20 nA) and voltage (0.01 V) on freestanding graphene, showing the height fluctuations over a time of 400 s. The slow scan direction is oriented horizontally. Scale bar, 0.02 nm. **b**, Height fluctuations as a function of time, taken by averaging the points in each vertical line of the lower STM image. For comparison, the same data was collected from a sample of gold. **c**, Actual tunneling current measured concurrently with the height fluctuations. **d**, Autocovariance of the height data, with an exponential decay constant of 8 s.

**Figure 3 | Fluctuations exhibiting periodic components. a**, Four line profiles taken in the same manner as those in Fig. 2b. All were acquired at a tip bias of 0.01 V, but the tunneling current was 3.00 nA for the top set and 5.00 nA for the bottom set. **b**, Associated $A(t)$ for each graphene curve in part **a**. This function shares the periodicity of the signal. Smoothed first derivatives (related to the linear response) were calculated using the Savitzky-Golay algorithm and are plotted as insets. **c**, Power spectral density for each graphene curve in part **a**, revealing the primary oscillation frequencies. **d**, The primary frequency as a function of $I^2$ for four images taken with a tip bias of 0.01 V. A trend line has been added to demonstrate the linear relationship

of the data. **e**, The height variance as a function of $I^{-4}$ for four images taken with a tip bias of 0.01 V.

**Figure 4 | The mirror buckling effect. a**, STM image of freestanding graphene, acquired in the same manner as that in Fig. 2a, but at a current of 4.00 nA and voltage of 0.1 V. Scale bar, 0.02 nm. **b**, Height profile extracted from the STM image and showing a jump of nearly 60 nm at $t = 100$ s, mimicking a critical transition. Corresponding data from a gold sample is provided for comparison. **c**, The measured tunneling current, which shows a large but unsaturated spike coinciding with the large, sudden displacement. **d**, $A(t)$ for the graphene data. **e**, The variance of the height-time graphs from images collected as the setpoint current was increased at a constant tip bias of 0.1 V. It is anomalously large when the mirror buckling effect occurs. **f**, Series of illustrations depicting the stages of mirror buckling for a hemispherical shell. Corresponding labels are applied to the data in **b**. A sudden inversion in the curvature occurs at the critical pressure, just before Stage 3.

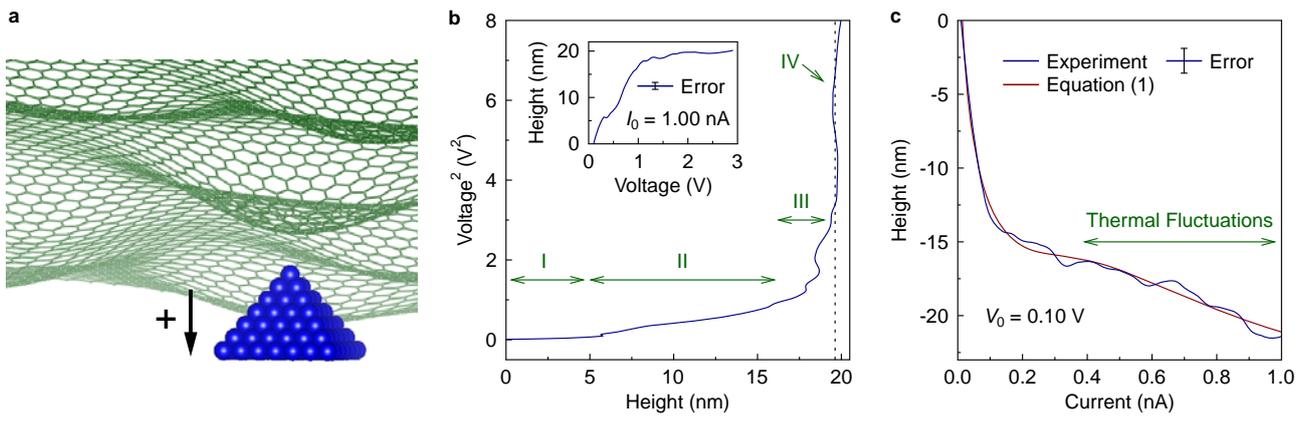

**Figure-1 (XU)**

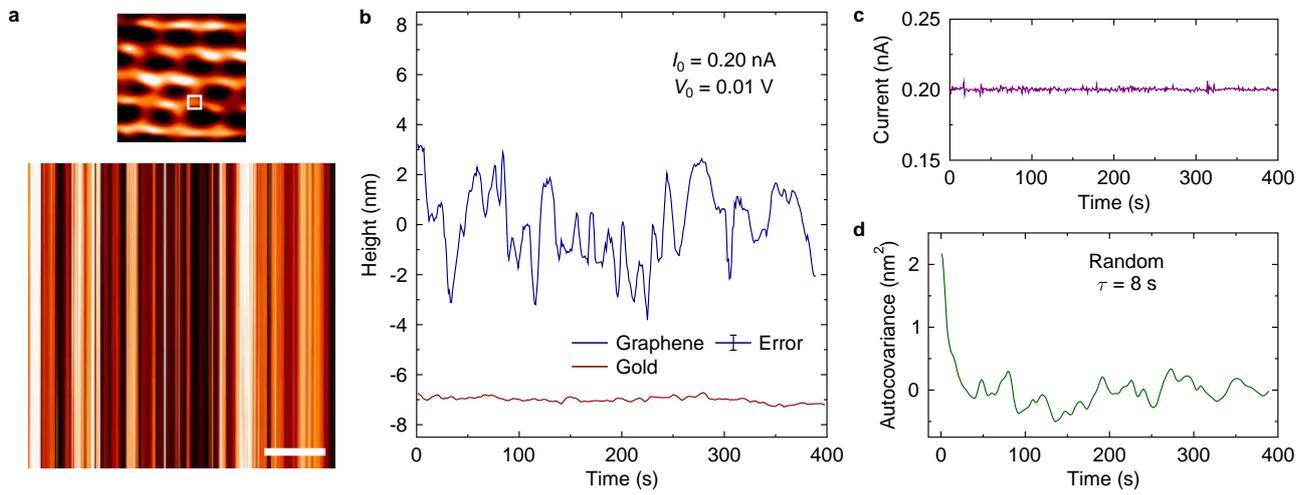

**Figure-2 (XU)**

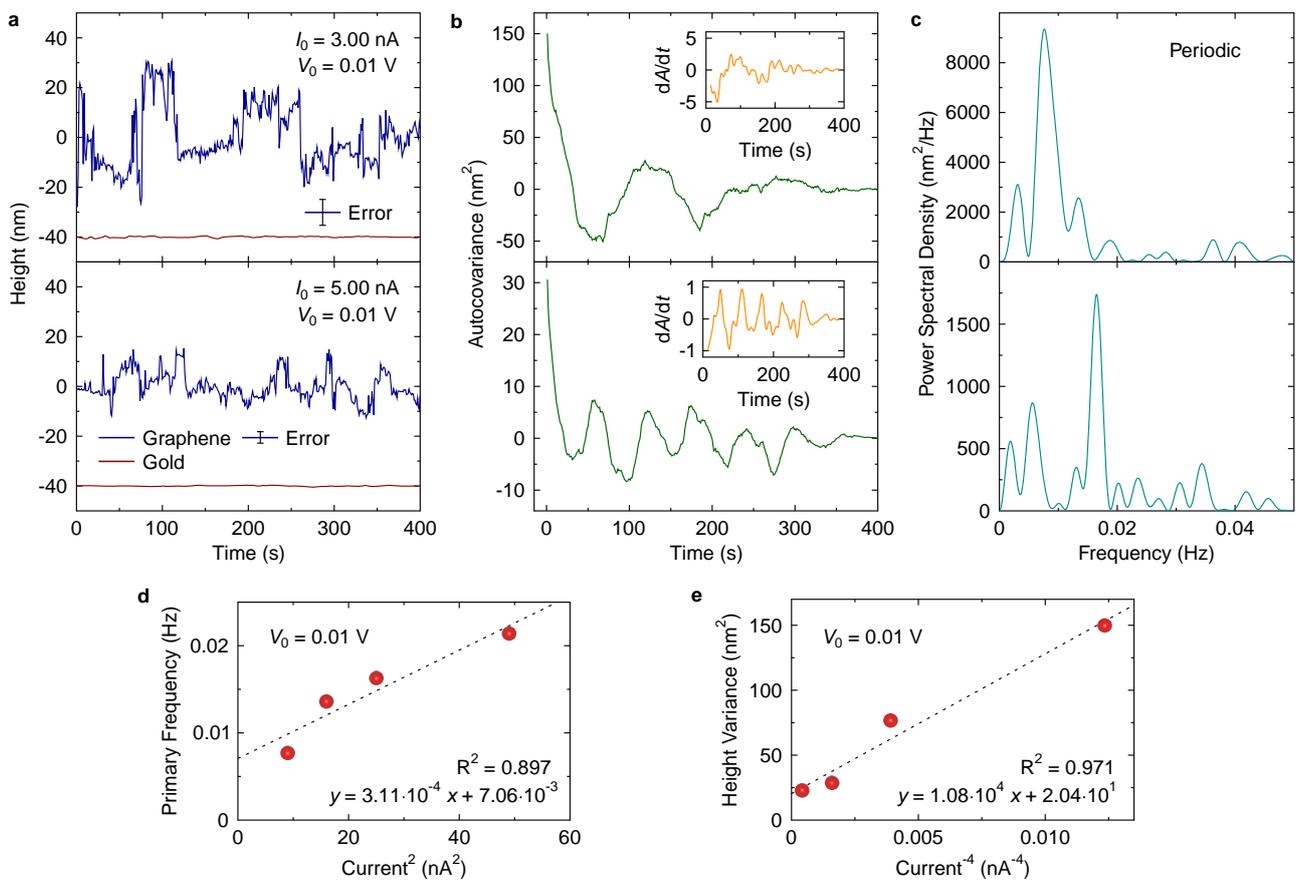

**Figure-3 (XU)**

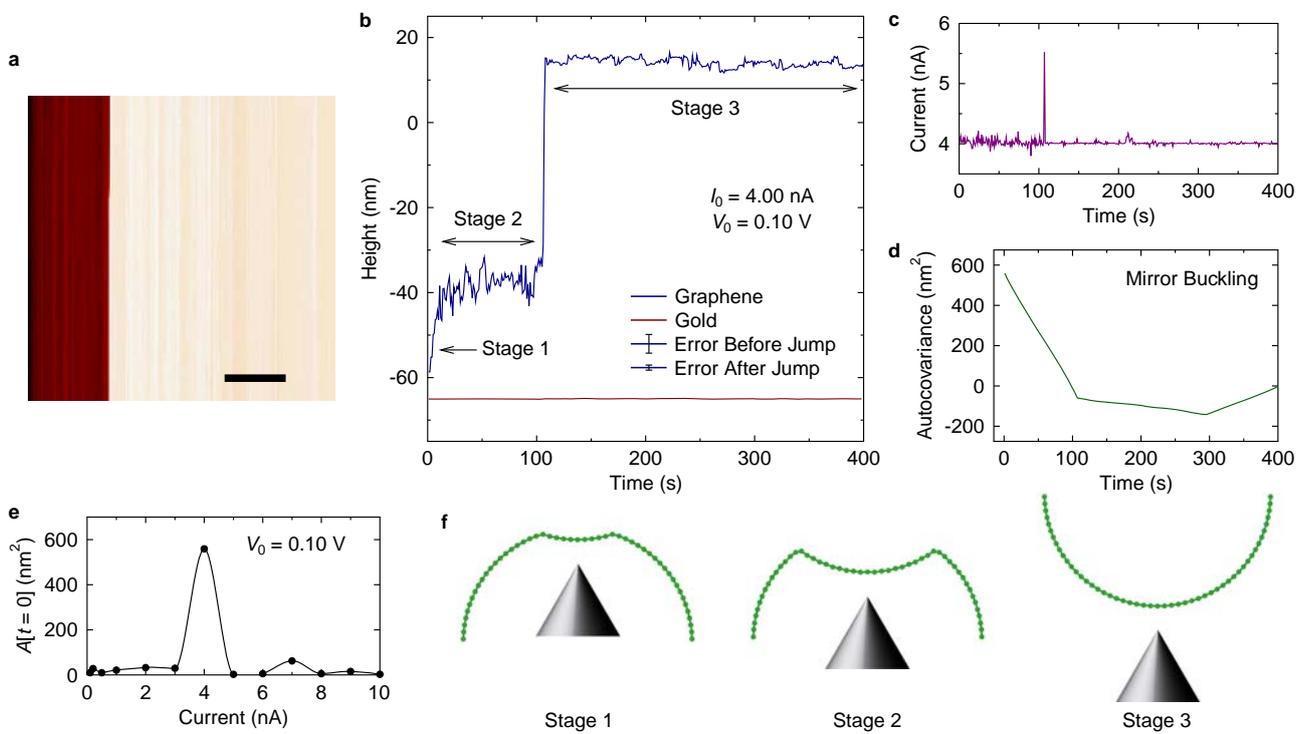

**Figure-4 (XU)**